\documentclass[a4paper,11pt]{article}
\usepackage{pos}
\usepackage{float}

\title{Updated constraints on WIMP dark matter annihilation by radio observations of M31 -- all annihilation channels}
\ShortTitle{Updated constraints on WIMP dark matter annihilation by radio observations of M31}

\author*{Andrei E. Egorov}

\affiliation{Nuclear Physics and Astrophysics Division, Lebedev Physical Institute,\\ 
	Leninskii prospect - 53, Moscow, Russia}

\emailAdd{aegorov@runbox.com}

\abstract{The present work derived the robust constraints on annihilating WIMP parameters utilizing new radio observations  of M31, as well as new studies of its DM distribution and other properties. The characteristics of emission due to DM annihilation were computed in the frame of 2D galactic model employing GALPROP code adapted specifically for M31. This enabled us to refine various inaccuracies of previous studies on the subject. DM constraints were obtained for all possible annihilation channels except $\chi\chi \rightarrow \gamma\gamma,\nu\bar{\nu}$. A wide variety of radio data was utilized in the frequency range $\approx$(0.1--10) GHz. As the result the thermal WIMP lighter than fiducially $\approx$(40--70) GeV was excluded in the case of light primary annihilation products $\chi\chi \rightarrow \tau^+\tau^-,\mu^+\mu^-,gg,c\overline{c},u\overline{u},d\overline{d},s\overline{s},e^+e^-,b\overline{b}$ (and an arbitrary combination of them). Heavier WIMP, which can annihilate to $W^+W^-,Z^0Z^0,t\overline{t},hh$; can not be probed at the level of thermal cross section, unless one assumes the optimistic cases of DM density and magnetic field distributions in M31. In summary, $m_x \gtrsim 40$ GeV represents the fiducial channel-independent mass limit for the thermal WIMP with the full uncertainty range estimated to be $\approx$(20--90) GeV.            The obtained exclusions are competitive to those from Fermi-LAT observations of dwarfs and AMS-02 measurements of antiprotons. Our constraints significantly restrict the opportunity to explain the gamma-ray outer halo of M31 by annihilating DM. And, finally, we questioned the possibility claimed in other studies to robustly constrain heavy thermal WIMP with $m_x > 100$ GeV by radio data on M31.}

\FullConference{
  *** 27th European Cosmic Ray Symposium - ECRS ***\\
  *** 25-29 July 2022 ***\\
  *** Nijmegen, the Netherlands ***
}

\begin{document}
\maketitle

\section{\label{sec:i}Introduction and motivation}
The physical nature of dark matter (DM) remains to be one of the biggest puzzles in modern physics and astronomy. Currently, we may outline three most popular candidates for the role of DM: weakly interacting massive particles (WIMPs), axionlike particles (ALPs) and sterile neutrinos (see e.g. \cite{2019arXiv191204727R} for a review). This work is dedicated to indirect searches of WIMPs, which traditionally have been the most anticipated. In our case we consider DM halo of the big neighbor galaxy M31 (Andromeda galaxy). Annihilating WIMPs in its halo would produce relativistic electrons and positrons ($e^\pm$), which in turn would generate synchrotron emission in the galactic magnetic field (MF) at radio frequencies. Hence, radio observations of the galaxy may infer some constraints on annihilating DM or even hints of its signal. Meanwhile, historically, the first attempt (known to the author) to constrain DM particle properties by radio observations was made in \cite{1992PhLB..294..221B}.

Traditionally the gamma-ray band has been the most promising for WIMP indirect searches: space-based gamma-ray telescopes, particularly Fermi-LAT, are able to probe the thermal WIMPs up to $m_x \approx 100$ GeV by observations of Milky Way (MW) dwarf satellites \cite{2020JCAP...02..012H}; and ground-based facilities, particularly CTA, will be able to probe the thermal WIMPs above this mass scale by observations of the Galactic halo \cite{2021JCAP...01..057A}. However, the mass range around $m_x \approx 100$ GeV is poorly reachable by both techniques. And here it is very relevant to employ alternative methodologies -- particularly, radio observations -- in order to complement the gamma-ray window and explore the whole WIMP mass range evenly.

M31 is one of the best targets in the sky for DM searches in radio due to its proximity, which allows detailed imaging and studies of the galactic medium. Also the central region of M31 is relatively faint in radio, which implies higher sensitivity to the potential DM signal due to lower hindering by usual astrophysical emissions. Being motivated by these advantages we started our work on the subject by \cite{2013PhRvD..88b3504E}. The current work refined and updated the former (and also the works by other authors) in a variety of aspects. All the details are thoroughly described in \cite{2022PhRvD.106b3023E} (sec. II there detailizes inaccuracies of the previous works on the subject). This letter provides a brief summary and generalization for all possible annihilation channels (in \cite{2022PhRvD.106b3023E} only popular $\chi\chi \rightarrow b\bar{b},\tau^+\tau^-$ channels were computed). 

At first we modeled theoretically the characteristics of anticipated emission due to DM annihilation, i.e. the emission intensity dependence on all the parameters. This modeling procedure is described in the next sec. \ref{sec:dm}. Then the relevant observational data were collected and compared with the theoretically calculated DM emission intensities in order to derive DM constraints of interest -- this is described in sec. \ref{sec:constr}. And sec. \ref{sec:summary} provides a summary and relation with the gamma-ray band.

\section{\label{sec:dm}Modeling the emission due to DM annihilation by GALPROP}

For this purpose one has to solve the transport equation for $e^\pm$ produced by DM annihilation (DM $e^\pm$ hereafter) and then to calculate the synchrotron emission from them assuming certain MF distribution. This task was fully solved by employment of GALPROP package \cite{GP} (v56, specifically adapted for M31) together with my addition \cite{github}, which precisely calculates DM source term for the transport equation. The former has the following form (in general case of an arbitrary combination of the annihilation channels):
\begin{equation}\label{eq:q}
q(R,E) = \frac{1}{2} \langle \sigma v \rangle \left(\frac{\rho(R)}{m_x}\right)^2 \xi(R) \sum_i BR_i \frac{dN_{e,i}}{dE}(E,m_x),	
\end{equation}
where $\langle \sigma v \rangle$ denotes WIMP annihilation cross section, $\rho(R)$ is DM density distribution, $m_x$ is WIMP mass, $\xi(R)$ is the DM annihilation rate boost factor due to substructures, $BR_i$ is the branching ratio to the $i$-th channel (summation goes over all channels and $\sum_i BR_i = 1$) and $\frac{dN_{e,i}}{dE}(E,m_x)$ represents the energy spectra of $e^\pm$ per annihilation. These spectra were taken from PPPC 4 DM ID resource \cite{PPPC}. The transport equation for DM $e^\pm$ was solved in 2D, which is a proper approximation for a spiral galaxy. All the previous works employed a deficient 1D (i.e. spherical symmetry) approximation.

Indeed there are significant uncertainties in the emission modeling procedure. The main sources of uncertainties are DM density and MF distributions. $e^\pm$ propagation parameters are uncertain too, but they play a secondary role. In order to quantify the uncertainties thoroughly I employed the quite traditional MIN-MED-MAX paradigm, when MIN and MAX model setups provide respectively the lowest and highest DM emission intensities and, hence, the weakest and strongest possible constraints. And MED setup represents some middle or "average expectation" scenario (but not necessarily the average intensity). MIN-MED-MAX models were built separately for DM density profile and MF distribution together with the propagation parameters (MF/prop. with "/" meaning "and") -- i.e. three independent density profiles and three independent MF/prop. configurations were tested. MF and propagation parameters were treated jointly in order to reduce a computational heaviness of the task taking into account a subdominant role of the propagation uncertainties. Thus, for each discrete WIMP mass and annihilation channel 3 DM density profiles $\times$ 3 MF/prop. configurations = 9 independent models were computed, providing a good coverage of the parameter space. 1161 separate GALPROP runs were made in order to explore all the relevant WIMP masses and annihilation channels.

In order to set MIN-MED-MAX DM density profiles I conducted a comprehensive analysis of various profile determinations available in the literature with especial emphasis to newer works, which emerged in recent years. Several papers on the subject agree that very cuspy or cored profiles are not favored for M31. Finally I chose Einasto density profiles from \cite{2018MNRAS.481.3210B} and \cite{2012A&A...546A...4T} for MIN and MAX cases respectively, and NFW profile from \cite{2018MNRAS.481.3210B} for MED. I refer readers to \cite{2022PhRvD.106b3023E} for the fully-detailed description. Here the description is very brief due to a paper length limit. Regarding the MF distribution model, it is 2D and comprises piecewise-linear dependence on the radius $r$ multiplied by the exponential dependence on the vertical coordinate $z$. MED and MAX MF models have additional central cusps, which were motivated by MF determination in MW central region \cite{2010Natur.463...65C} and possible similarity between two galaxies. MF strength in the central region of M31 has the biggest importance for our work, since the potential DM signal is very concentrated around the galactic center. I set the central field values based mainly on the results of \cite{1998IAUS..184..351H,2010Natur.463...65C}. These values are the following: for MIN -- 14 $\mu$G, for MED -- 50 $\mu$G and for MAX -- 100 $\mu$G. $e^\pm$ propagation parameters were modeled based on CR diffusion studies for both M31 and MW. One important peculiarity in M31 nucleus is that $e^\pm$ there have much faster cooling rate through the inverse Compton scattering losses due to denser (by $\approx$2 times) radiation field in comparison with MW nucleus. And the free-free absorption of the synchrotron emission of interest (at $\nu \gtrsim$ 0.1 GHz) was proved to be negligible.

Putting together all the ingredients described above, DM emission maps and spectra were computed. Their examples and parameter dependencies can be seen in \cite[sec. IV]{2022PhRvD.106b3023E}.

\section{\label{sec:constr}Derivation of DM constraints from various radio data}

M31 images at eight frequencies in the range $\approx$(0.1--10) GHz were employed for the derivation of constraints. These maps are mainly clean of the projected discrete sources. On some images even the diffuse thermal emission was subtracted. The main role (by constraining power) belongs to very new and sensitive data from LOFAR Two-meter Sky Survey (LoTSS) \cite{2022A&A...659A...1S}. Regarding a choice of the region of interest (ROI) on the sky for derivation of constraints, the central (i.e. bulge) region with $R \lesssim 3$ kpc was employed, since it comprises a vast majority of the emission flux due to DM. Outside of this radius DM emission intensity becomes too low w.r.t. the usual astrophysical backgrounds. Intrinsically this bulge region was divided to concentric annuli individually at each frequency according to the corresponding image resolution. And the observed emission intensity in each annulus at each frequency entered the likelihood function individually and independently. Thus the latter has 33 components/multipliers in total. An important aspect in our model is an absence of any specific assumptions about the intensity of standard astrophysical emission, i.e. the flat independent priors were set for it at all frequencies. This essentially means that the upper limit of the total observed intensity defines the limit for DM emission. This makes our DM constraints quite robust and model-independent. And our likelihood function includes also the systematic uncertainties for the observed intensity besides the statistical uncertainties in the form of map noise. 

\begin{figure}[H]
	\includegraphics[width=0.497\linewidth]{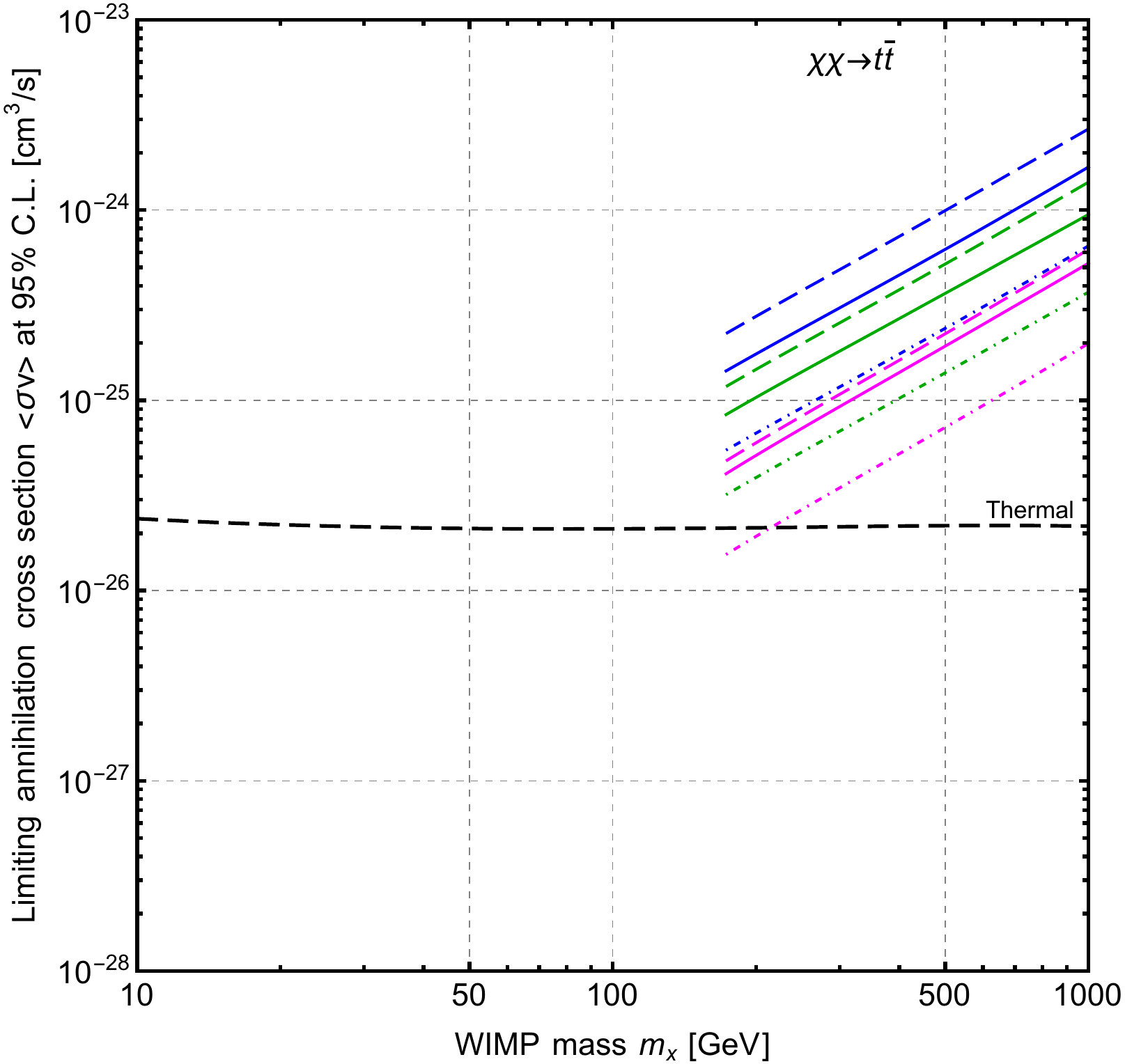}
	\includegraphics[width=0.497\linewidth]{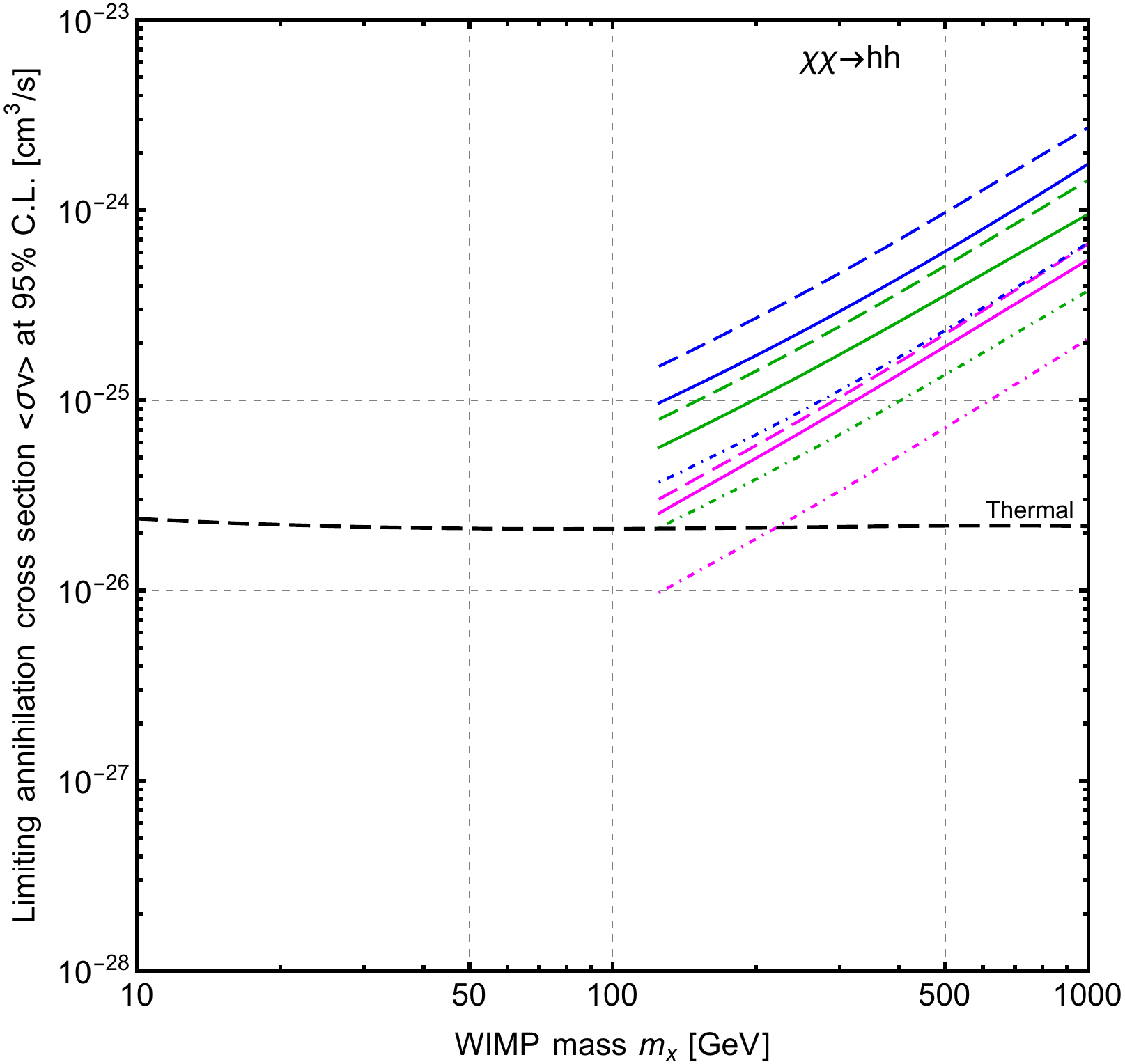}
	\caption{\label{fig:sv}The limits on annihilation cross section vs. WIMP mass for various annihilation channels marked at each panel ($q \equiv u,d,s$). Each panel shows all nine computed configurations of DM density profiles and MF/prop. The (almost) horizontal dashed line shows the thermal relic cross section value (taken from \cite{2020JCAP...08..011S}).}
\end{figure}

After construction of the likelihood, which is a function of DM parameters and the observed intensities with their uncertainties, it was marginalized over nuisance parameters in order to get the probability density for DM annihilation cross section. DM emission intensity is linearly proportional to the cross section, if other DM parameters are fixed. Thus the limiting $\langle \sigma v \rangle$ values were numerically calculated for each of 1161 DM models. The results are presented in fig. \ref{fig:sv}. Overall, we do not see any drastic differences between the exclusion bands for various channels. A general trend is that the exclusion lines for the hadronic channels have smaller slope (i.e. $\langle \sigma v \rangle_{\text{lim}}$ increases with the mass slower) and wider uncertainty range w.r.t. the leptonic channels. Considering heavier thermal WIMP annihilating to $W^+W^-,Z^0Z^0,t\overline{t},hh$, it might manifest itself only in the case of optimistic configurations of DM density and MF/prop. A typical width of the whole uncertainty bands in fig. \ref{fig:sv} in the vertical direction (i.e., over $\langle \sigma v \rangle$) slightly exceeds one order of magnitude.

\begin{figure}[H]
	\includegraphics[width=1\linewidth]{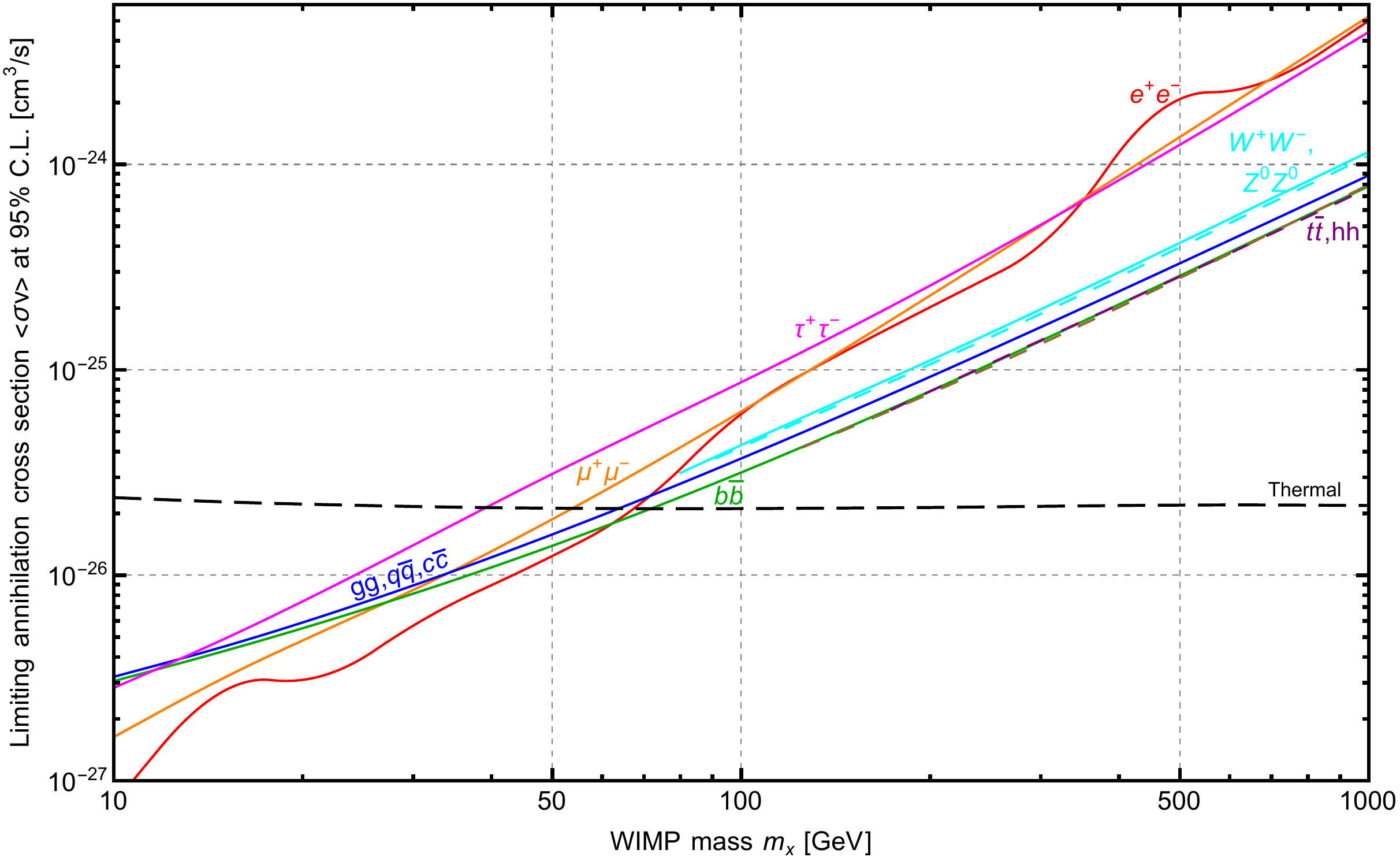}
	\caption{\label{fig:all}The fiducial or effective average exclusions for all the channels together. These exclusions represent the geometric averages of all nine DM density and MF/prop. models.}
\end{figure}

We also naturally would like to have some representative constraints, which are averaged over the uncertainties. For this purpose the geometric mean of all nine DM density and MF/prop. models was calculated for each channel as the fiducial or effective average constraint. The result is shown in fig. \ref{fig:all} for all the channels. We see that the exclusions for all channels form a relatively narrow band -- narrower than one order of magnitude over the cross section at any WIMP mass. For $m_x \gtrsim 60$ GeV the exclusions for all channels are mainly confined between those for $b\overline{b}$ and $\tau^+\tau^-$. And $\tau^+\tau^-$ appears to be the least constrained channel over almost the whole mass range considered. Table \ref{tab} lists WIMP mass limits and their uncertainty ranges for the thermal cross section. The "..." symbol in the table means that the corresponding exclusion curve does not reach the thermal level at any WIMP mass; i.e. our setup is not sensitive enough to probe the corresponding channel, and any WIMP mass is allowed.
\begin{figure}[H]
	\includegraphics[width=0.497\linewidth]{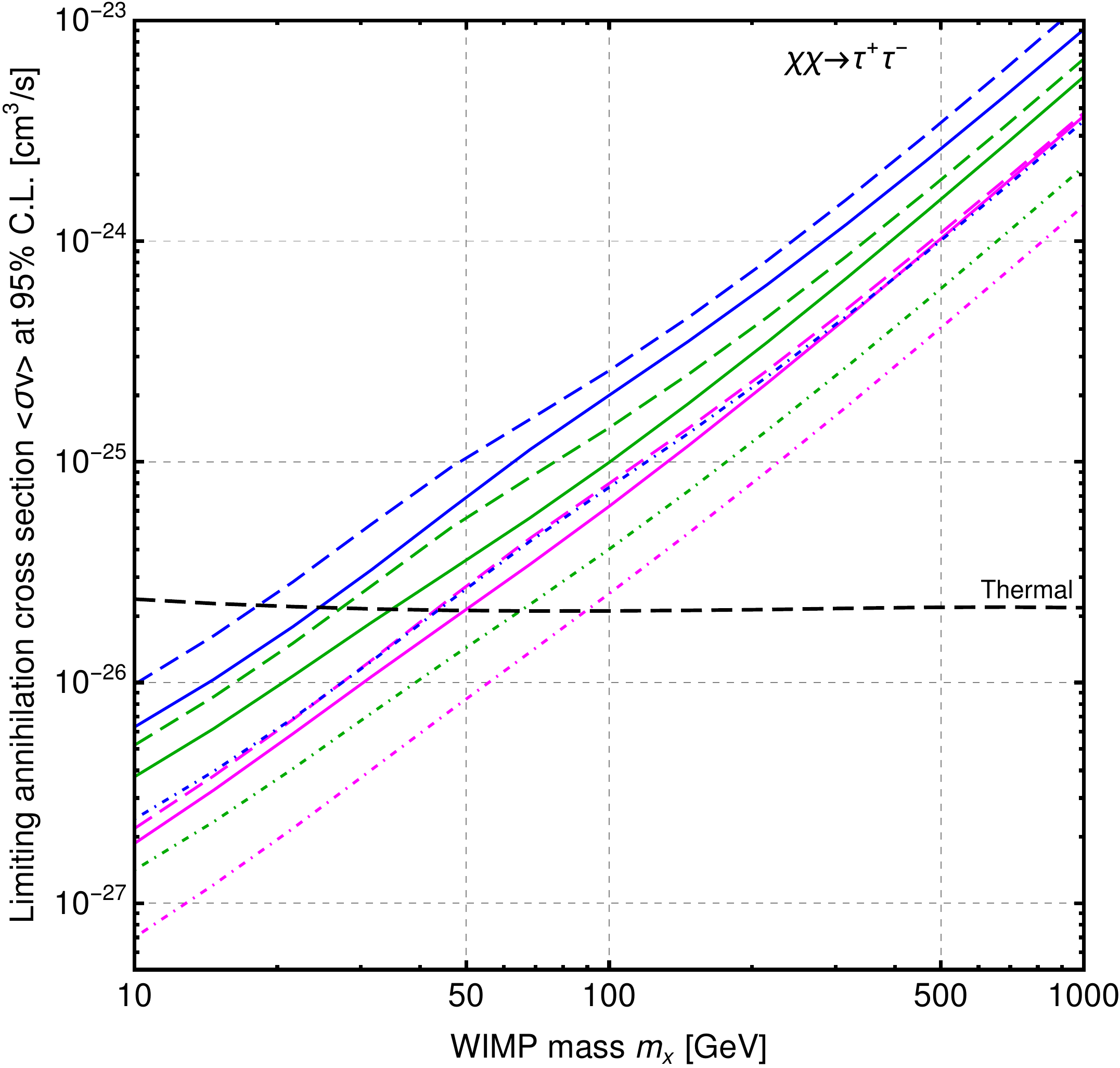}
	\includegraphics[width=0.497\linewidth]{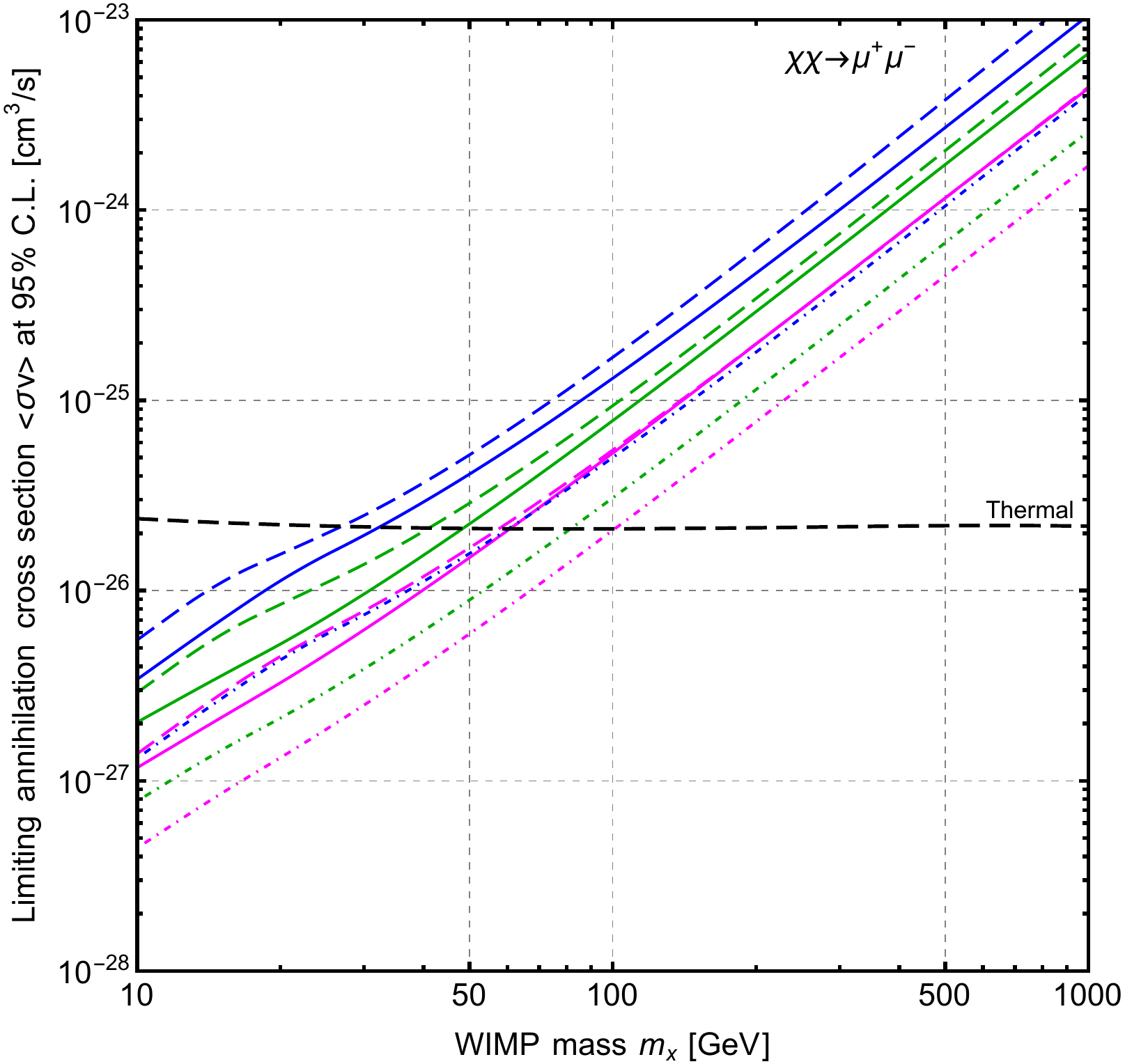}
	\includegraphics[width=0.497\linewidth]{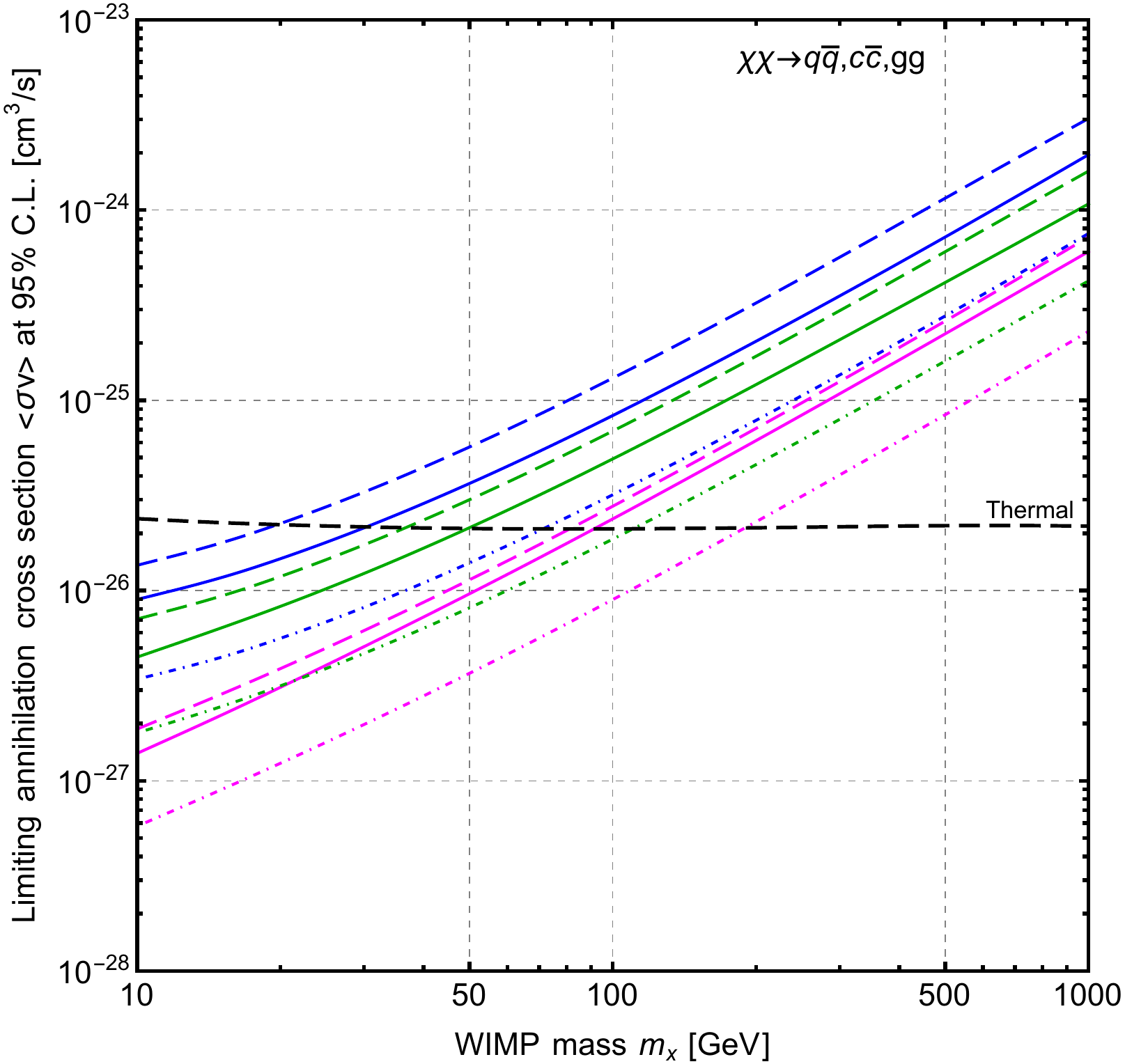}
	\includegraphics[width=0.497\linewidth]{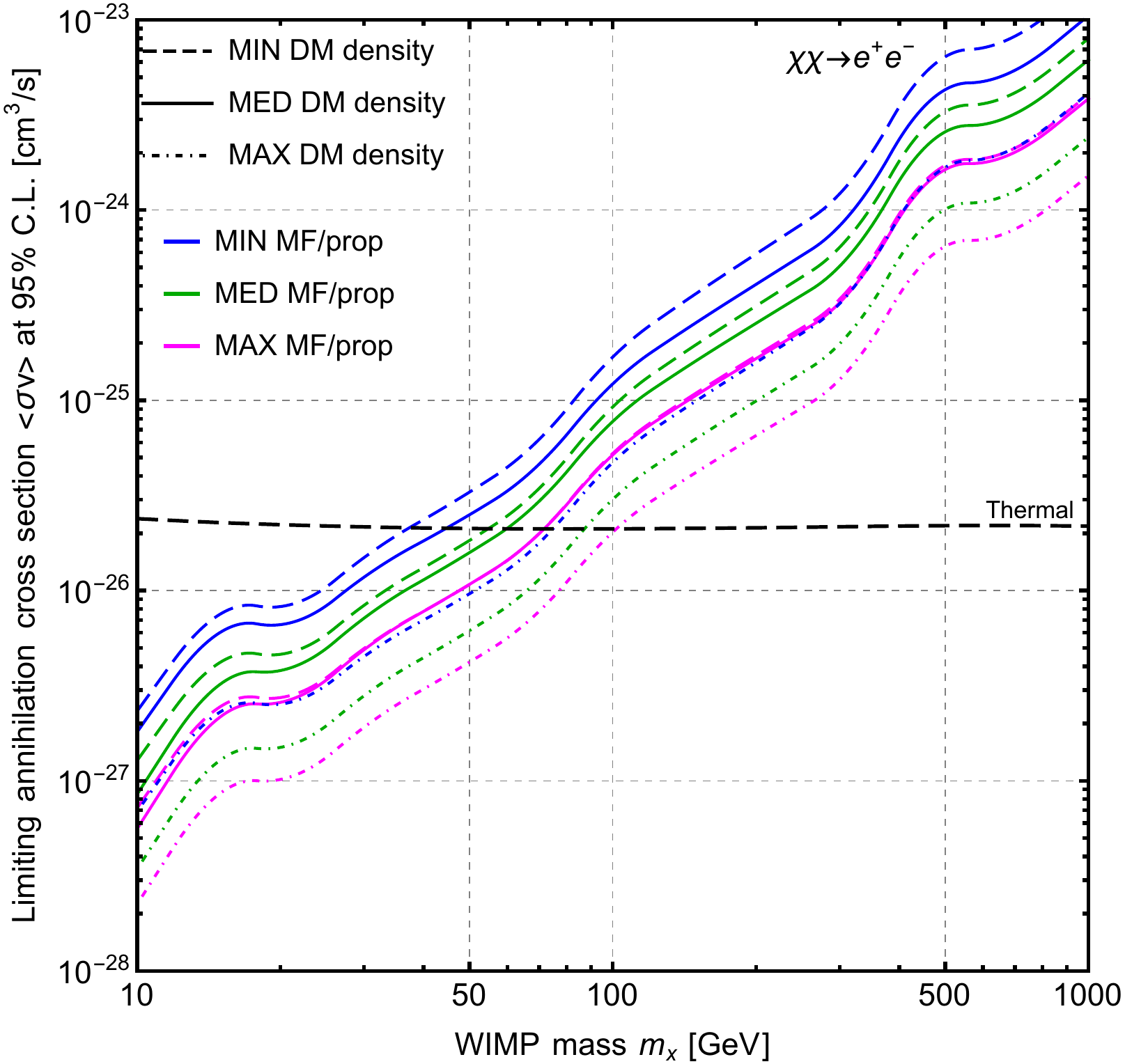}
	\includegraphics[width=0.497\linewidth]{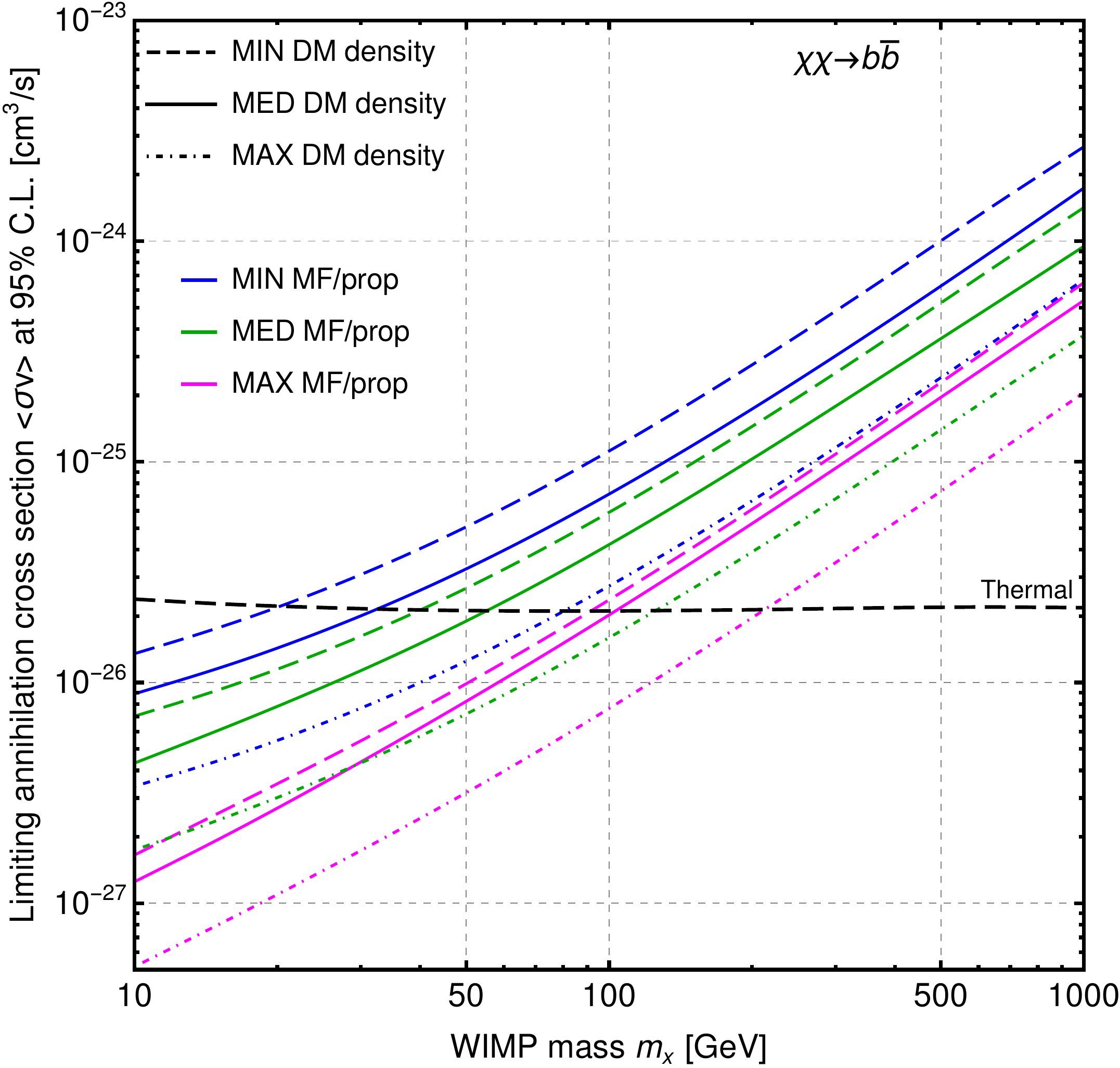}
	\includegraphics[width=0.497\linewidth]{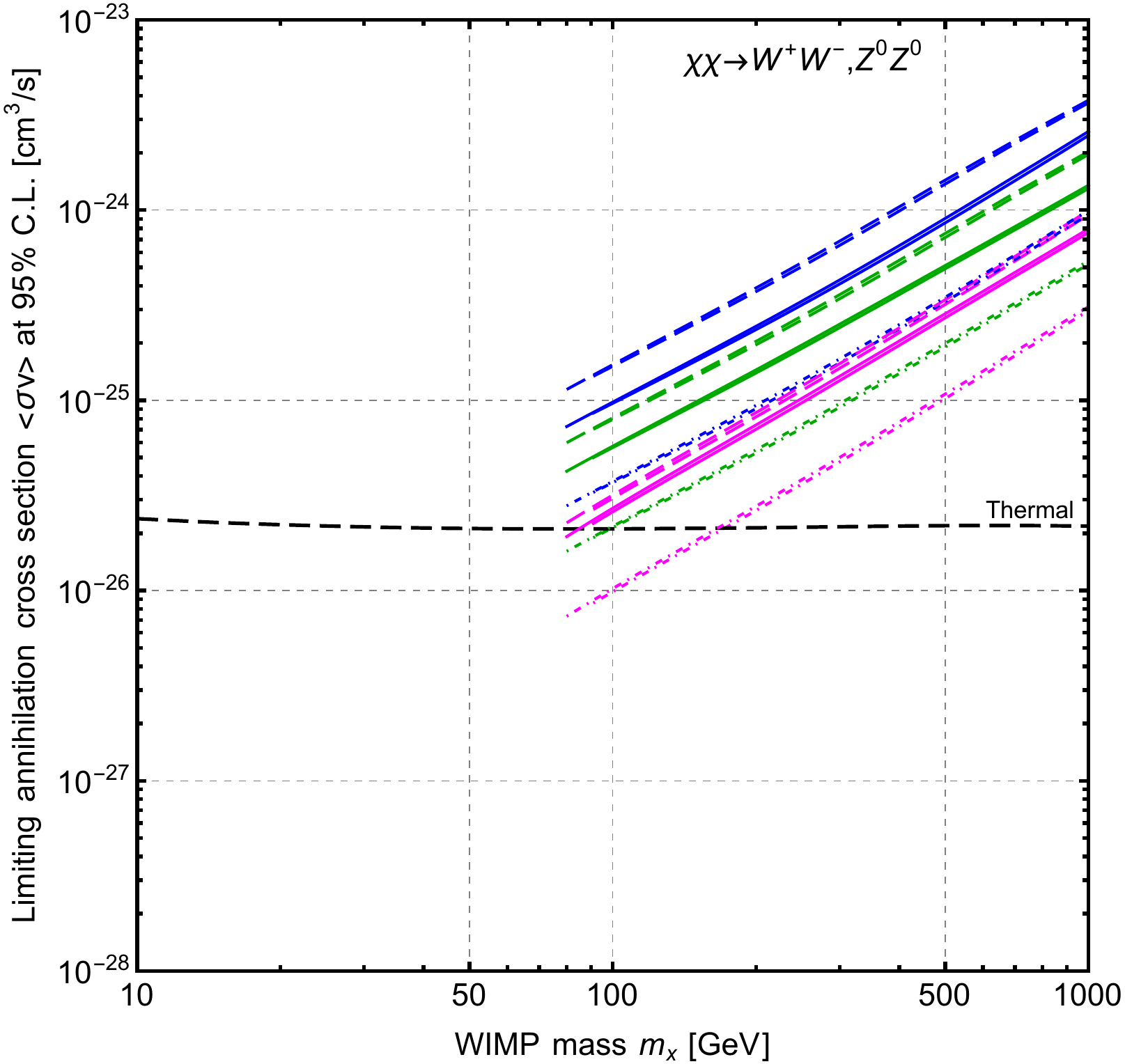}
\end{figure}

Another natural opportunity, which should be taken into account, is the annihilation through several channels simultaneously. It is easy to prove algebraically, that in the case of an arbitrary mixture of channels the limit for cross section is confined between the limits for the least and most constrained channels in the mixture, when the latter channels are considered alone. In general, WIMPs may also annihilate directly to photons or neutrinos, although these channels are highly suppressed in many models \cite{PPPC}. $\chi\chi \rightarrow \gamma\gamma$ channel is probed effectively by gamma-ray observations \cite{2020JCAP...11..049E}, which severely exclude the thermal WIMP, up to $m_x \sim 10$ TeV! 

\begin{table}[t]
	\caption{\label{tab}The obtained lower limits on the thermal WIMP mass (in GeV) and their uncertainty ranges.}
	\vspace{-0.3cm}
	\centering
		\begin{tabular}{cccc}
			\hline \hline
			&             & Range due to   & Range due to \\
			& Fiducial    & DM density     & DM density \\
			Annihilation & (geometric) & uncertainties  & and MF/prop. \\		
			channel      & average     & (MED MF/prop.) & uncertainties \\ 
			\hline
			$\chi\chi \rightarrow \tau^+\tau^-$ & \textbf{39} & 27--65 & 18--89 \\
			$\chi\chi \rightarrow \mu^+\mu^-$ & \textbf{54} & 41--81 & 27--100 \\
			$\chi\chi \rightarrow gg$ & \textbf{62} & 36--110 & 19--190 \\
			$\chi\chi \rightarrow c\overline{c}$ & \textbf{63} & 36--110 & 19--190 \\
			$\chi\chi \rightarrow u\overline{u},d\overline{d},s\overline{s}$ & \textbf{64} & 37--110 & 20--190 \\
			$\chi\chi \rightarrow e^+e^-$ & \textbf{67} & 54--89 & 37--100 \\
			$\chi\chi \rightarrow b\overline{b}$ & \textbf{72} & 40--120 & 20--210 \\
			$\chi\chi \rightarrow W^+W^-$ & \textbf{...} & ...--98 & ...--170 \\
			$\chi\chi \rightarrow Z^0Z^0$ & \textbf{...} & ...--99 & ...--170 \\
			$\chi\chi \rightarrow t\overline{t}$ & \textbf{...} & ... & ...--210 \\
			$\chi\chi \rightarrow hh$ & \textbf{...} & ... & ...--220 \\
			\hline \hline
		\end{tabular}		
\end{table}


\section{\label{sec:summary}Conclusions and discussion}

The robust WIMP annihilation constraints were derived here based on radio observations of the central (bulge) region of M31 in the wide frequency range $\approx$(0.1--10) GHz. Our model does not assume anything specific about the intensities of usual astrophysical emissions in M31. Table \ref{tab} summarizes the thermal WIMP mass lower limits for all annihilation channels. Thus, we may finally conclude that $m_x \gtrsim 20$ GeV represents the \textit{hard and channel-independent} limit for the thermal WIMP; i.e. the former holds under \textit{any} reasonable assumptions about DM density and MF distributions, and propagation parameters. $m_x \gtrsim 40$ GeV is the fiducial (effective average over the uncertainties) limit, and $m_x \gtrsim 90$ GeV is the most optimistic limit.

It was reported recently in \cite{2021PhRvD.103b3027K}, that M31 possess the extended gamma-ray outer halo, which may originate from annihilating WIMPs with $m_x \approx (45-72)$ GeV ($\chi\chi \rightarrow b\bar{b}$ channel). Our limits severely constrain this interpretation. The latter requires $\langle \sigma v \rangle \sim (10^{-26}-10^{-23})$ cm$^3$/s depending on the assumed DM density distribution. One can see from the corresponding panel of fig. \ref{fig:sv} above, that such high cross section values are majorly excluded even for MIN configuration of MF/prop.

\bibliographystyle{JHEP}
\bibliography{../../../universal}

\end{document}